\documentclass[10pt]{iopart}
\usepackage[latin9]{inputenc}
\usepackage{color}
\expandafter\let\csname equation*\endcsname\relax
\expandafter\let\csname endequation*\endcsname\relax
\usepackage{amsmath}
\usepackage{amssymb}
\usepackage{stmaryrd}
\usepackage{graphicx}

\makeatletter

\providecommand{\tabularnewline}{\\}

\usepackage{iopams}

\usepackage{color}
\usepackage{textcomp}
\usepackage{wasysym}
\usepackage{epstopdf}
\usepackage{caption}

\makeatother

\begin{document}

\title{General Lossless Spatial Polarization Transformations}

\author{Alicia Sit,$^{1}$ Lambert Giner,$^{1,*}$ Ebrahim Karimi,$^{1,2}$
and Jeff S. Lundeen$^{1}$}

\address{$^{1}$ Physics Department, Centre for Research in Photonics, University
of Ottawa, Advanced Research Complex, 25 Templeton, Ottawa ON Canada,
K1N 6N5}

\address{$^{2}$ Department of Physics, Institute for Advanced Studies in
Basic Sciences, 45137-66731 Zanjan, Iran}

\ead{$^{*}$lginer@uottawa.ca}
\begin{abstract}
Liquid crystals allow for the real-time control of the polarization
of light. We describe and provide some experimental examples of the
types of general polarization transformations, including universal
polarization transformations, that can be accomplished with liquid
crystals in tandem with fixed waveplates. Implementing these transformations
with an array of liquid crystals, e.g., a spatial light modulator,
allows for the manipulation of the polarization across a beam's transverse
plane. We outline applications of such general spatial polarization
transformations in the generation of exotic types of vector polarized
beams, a polarization magnifier, and the correction of polarization
aberrations in light fields.
\end{abstract}

Keyword : Polarization, Liquid-crystal devices, Quantum information
and processing 

\maketitle


\section{Introduction}

For the last century, polarization manipulation has mostly been conducted
using birefringent crystals, known as waveplates, by physically rotating
them about a light beam's propagation axis~\cite{Bhandari1989,Simon1990,Damask2004,DeZela2012,Born:80}.
However, waveplates manipulate the polarization uniformly across a
beam's transverse profile; that is, they do not allow for spatially
varying polarization manipulation. A cell of uniformly aligned liquid
crystals acts as a waveplate with a fixed orientation optical axis
and a voltage-controlled variable birefringence. When arranged in
an array, such as in a liquid-crystal spatial light modulators (LC-SLM),
these devices can spatially tailor the polarization distribution of
light by individually controlling the voltage across each cell.

LC-SLMs are widely used for dynamic generation of optical beams possessing
particular intensity and phase profile~\cite{Clark:16,Bolduc:13,Arrizon:07}.
In the past fifteen years, their inherent birefringence has been used
to produce arbitrary spatially polarized beams. However, the schemes
to do so are inherently lossy; they rely on spatial and polarization
filtering~\cite{Neil2002,Maurer2007,Wang2007,Franke-Arnold2007,Chen2011,Moreno2012,Han2013,Waller2013,Maluenda2013,Guo2014,Yu2015}.
While with classical optics and photonic communications this signal
loss is undesirable, in quantum optics it can completely destroy the
quantum nature of the light being acted on \cite{Walls2008}.

Consequently, the polarization transformation schemes that we present,
like the schemes in Refs. \cite{Davis2000,Eriksen2001,Kenny2012,Estevez2015,Zheng2015,Galvez:12},
do not require optical loss in order to function. Whereas all of these
references focus solely on generating spatially-varying polarized
states of light, we additionally investigate the implementation of
general spatial polarization transformations. Implementing these would
enable unprecedented control over spatially-varying polarization distributions.
Such control could have applications in studying the dynamics and
topologies of polarization vortices~\cite{Freund:06,Cardano:13,Khajavi:16,Otte:16,Pal:16},
creating exotic polarization topologies in beams, e.g., a Möbius strip
in polarization~\cite{Bauer2015}, creating novel optical traps for
biology and atomic physics~\cite{Franke-Arnold2007,Dholakia2011},
for optically guiding and pumping microfluids~\cite{Leach2006,Delville2009},
and for micro-machining~\cite{Schadt1992,Ambrosio:12,Toyoda:13}.
In the quantum realm, spatially polarized states can be used to multiply
communication bandwidth through superdense coding \cite{Milione2013,Li2015},
perform tests of fundamental quantum physics~\cite{Karimi2010,Karimi:14},
implement quantum key distribution~\cite{Vallone2014,Sit:16,Ndagano:17},
and for building measurement devices with quantum-enhanced sensitivities~\cite{Toppel2014,Berg:15}.

We limit ourselves to light fields that are perfectly polarized at
each spatial point and to transformations that maintain this perfect
degree of polarization. Moreover, the transformations should not involve
loss, at least in principle. These are known as unitary transformations
and are mathematically enacted by Jones Matrices \cite{Jones1947}.
In turn, these are equivalent to rotations in the Poincaré sphere,
as we will review in Section 2 and use throughout this paper. In Section~\ref{sec:general_pol_transfo},
we will discuss three different kinds of universal unitaries that
can be created by combining voltage-controlled liquid crystal cells
and fixed waveplates: 1. Variable phase retardation of a fixed but
arbitrary polarization. 2. Transformation from an arbitrary polarization
state to another arbitrary polarization state. 3. Variable retardation
of a variable arbitrary polarization state. At the expense of requiring
increasing numbers of waveplates and liquid crystal devices, going
from the first to last, these transformations increase in generality.
The latter is the most general unitary possible for polarization transformations. 

The mathematical theory underlying these transformations has not been
previously presented. The paper provides explicit formulae and the
underlying definitions and conventions that are needed to implement
these general polarization transformations in practice. Moreover,
the distinct goals of arbitrary polarization generation and arbitrary
transformations are often conflated. We clarify the fundamental differences
between them and show that have different requirements and constraints.

\section{Background theory}


\subsection{Waveplates and liquid crystal cells}

In this section, we describe the transformation of polarization by
birefringent media in terms of rotations in the Poincaré sphere. Since
many mutually inconsistent polarization conventions exist in the literature,
we give a brief introduction and review of polarization transformations
in Appendix A, which sets the conventions and notation used in this
paper. Light passing through birefringent media, such as a waveplate,
gains a phase between the electric field component along the media's
optical axis and the orthogonal component. This phase is known as
the ``retardance,'' $\Delta$. On the Poincaré sphere, the action
of a waveplate corresponds to a rotation of an input polarization
$\mathbf{\hat{s}}$ by $\Delta$ about a rotation axis $\mathbf{\hat{k}}\left(2\Phi,0\right)$,
i.e., one lying in the $\mathbf{\hat{s}}_{1}$$\mathbf{\hat{s}}_{2}$
plane at $2\Phi$ from the positive $\mathbf{\hat{s}}_{1}$ axis.
Here, $\Phi$ is the angle in the laboratory between the fast axis
and horizontal, $\hat{\mathbf{x}}$, with increasing angle defined
to be towards $\hat{\mathbf{y}}$. Accordingly, waveplates, such as
half-wave plates ($\Delta=\pi$, HWP) and quarter-wave plates ($\Delta=\pi/2$,
QWP), can be described by a rotation matrix. More generally, a birefringent
waveplate with arbitrary retardance $\Delta$ (e.g., an electro-optic
modulator (EOM) or a nematic-phase parallel-aligned liquid crystal)
will have the rotation matrix \cite{Koks2006},

\begin{equation}
\mathbf{R}_{\mathbf{k}\left(2\Phi,0\right)}(\Delta)=\left[\begin{array}{ccc}
\sin^{2}2\Phi\cos\Delta+\cos^{2}2\Phi & \sin^{2}\left(\frac{\Delta}{2}\right)\sin4\Phi & \sin2\Phi\sin\Delta\\
\sin^{2}\left(\frac{\Delta}{2}\right)\sin4\Phi & \cos^{2}2\Phi\cos\Delta+\sin^{2}2\Phi & -\cos2\Phi\sin\Delta\\
-\sin2\Phi\sin\Delta & \cos2\Phi\sin\Delta & \cos\Delta
\end{array}\right].\label{eq:wp}
\end{equation}

Since we will be using HWP and QWPs frequently, we define $\mathbf{HWP}\llbracket\Phi\rrbracket\equiv\mathbf{R}_{\mathbf{k}\left(2\Phi,0\right)}(\pi)$
and $\mathbf{QWP}\llbracket\Phi\rrbracket\equiv\mathbf{R}_{\mathbf{k}\left(2\Phi,0\right)}(\pi/2)$.
To avoid any confusion between the reference frames the angles are
defined in, an angle expressed the laboratory frame is written using
$\llbracket\cdot\rrbracket$.



In the Poincaré sphere, an arbitrary general polarization transformation
$\mathbf{R}_{\mathbf{k}}\left(\xi\right)$ consists of a rotation
about an arbitrary axis $\mathbf{\hat{k}}$ by angle $\xi$ (i.e.,
Eq.~(\ref{eq:E-R})). One can implement this with half- and quarter-wave
plates cascaded in the following sequence: 
\begin{equation}
\mathbf{R}_{\mathbf{k}}\left(\xi\right)=\mathbf{QWP}\textrm{\ensuremath{\llbracket\Phi_{3}\rrbracket}}\mathbf{HWP}\textrm{\ensuremath{\llbracket\Phi_{2}\rrbracket}}\mathbf{QWP}\textrm{\ensuremath{\llbracket\Phi_{1}\rrbracket}}.\label{eq:simon}
\end{equation}
Ordered first to last, the waveplates the light passes through respectively
correspond to the elements in Eq. (\ref{eq:simon}) from right to
left. By an appropriate rotation of each of the three waveplates in
the laboratory, any unitary polarization transformation can be created
\cite{Simon1990}. However, this mechanical rotation will prohibit
fast changes of the unitary polarization transformation.

Now suppose we had a birefringent optical element with retardance
$\Delta$ and rotation axis $\mathbf{\hat{k}}_{A}$, and we wished
to convert $\mathbf{\hat{k}}_{A}$ to be $\mathbf{\hat{k}}_{B}$,
as in Fig. 2(a). This could be done by sandwiching the optical element
between two sequences of Eq.~(\ref{eq:simon}),

\begin{eqnarray}
\mathbf{R}_{\mathbf{k}_{B}}(\Delta) & = & \mathbf{QWP}\left\llbracket \Phi_{1}+90^{\circ}\right\rrbracket \mathbf{HWP}\left\llbracket \Phi_{2}+90^{\circ}\right\rrbracket \mathbf{QWP}\left\llbracket \Phi_{3}+90^{\circ}\right\rrbracket \nonumber \\
 & \cdot & \mathbf{R}_{\mathbf{k}_{A}}(\Delta)\mathbf{QWP}\textrm{\ensuremath{\left\llbracket \Phi_{3}\right\rrbracket }}\mathbf{HWP}\textrm{\ensuremath{\left\llbracket \Phi_{2}\right\rrbracket }}\mathbf{QWP}\textrm{\ensuremath{\left\llbracket \Phi_{1}\right\rrbracket }},\label{eq:gen_rot_axis}
\end{eqnarray}
where $\mathbf{R}_{\mathbf{k}_{A}}(\Delta)$ is the corresponding
rotation matrix of the birefringent optical element, and again, light
passes through the elements from right to left. The first sequence
of quarter- and half-wave plates rotates the polarization state $\mathbf{\hat{s}}_{B}$
(positioned at $\mathbf{\hat{k}}_{B}$ on the sphere) to $\mathbf{\hat{s}}_{A}$
(positioned at $\mathbf{\hat{k}}_{A}$). The second sequence applies
the reverse rotation, rotating it back to $\mathbf{\hat{s}}_{B}$.
In essence, this is a change of basis such that $\mathbf{\hat{s}}_{B}$,
a polarization eigenstate of $\mathbf{R}_{\mathbf{k}_{B}}(\Delta)$,
is unaffected by the waveplates and optical elements \textemdash{}
apart from a global phase \textemdash{} and all other polarization
states undergo a rotation by $\Delta$ about $\mathbf{\hat{k}}_{B}$.

\begin{figure}[ht!]
\centering\includegraphics[width=13cm]{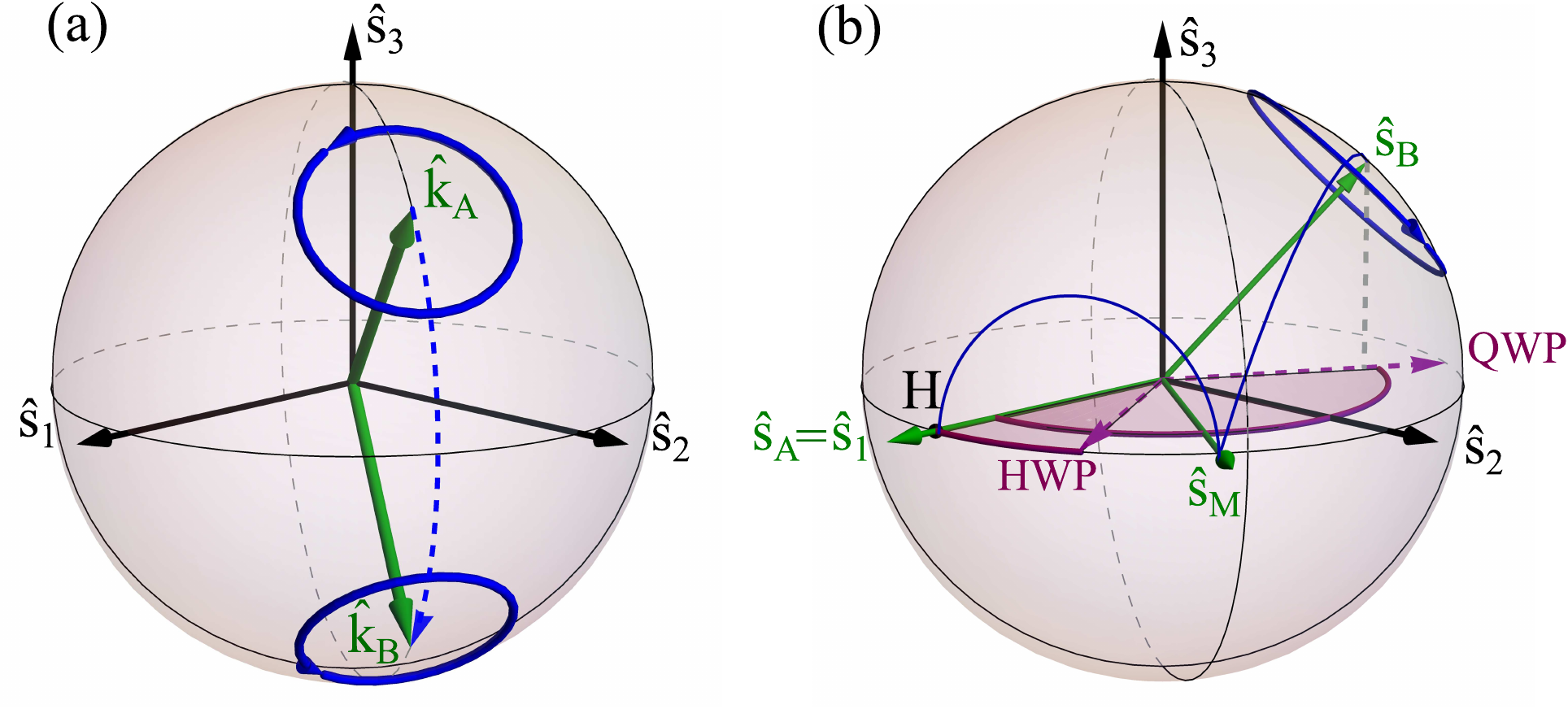} \protect\caption{(a) Representation on the Poincaré Sphere of the transformation of
an arbitrary rotation axis $\mathbf{\hat{k}}_{A}$ into another one
$\mathbf{\hat{k}}_{B}$. To perform such a transformation, a sequence
of three waveplates (QWP$\rightarrow$HWP$\rightarrow$QWP) converts
$\mathbf{\hat{k}}_{A}$ into $\mathbf{\hat{k}}_{B}$. (b) For a fixed
liquid crystal to perform a rotation about an arbitrary axis $\mathbf{\hat{k}}_{B}$,
one needs to convert the crystal's rotation axis, $\mathbf{\hat{k}}_{A}$
to $\mathbf{\hat{k}}_{B}$ (we take $\mathbf{\hat{k}}_{A}=\mathbf{\hat{s}}_{1}$).
Consider the passage of state $\mathbf{\hat{s}}_{B}=\mathbf{\hat{k}}_{B}$
through the sequence of waveplates. First a QWP removes the ellipticity
of $\mathbf{\hat{s}}_{B}$, transforming it into $\mathbf{\hat{s}}_{M}$,
which lies on the equator of the sphere. Then, a HWP rotates $\mathbf{\hat{s}}_{M}$
to $\mathbf{\hat{s}}_{A}=\mathbf{\hat{s}}_{1}=\mathbf{\hat{k}}_{A}$,
the eigenbasis of the LC-SLM. A general polarization state will be
rotated by $\Delta(x,y)$ around $\mathbf{\hat{s}}_{1}$, here. To
finish, $\mathbf{\hat{s}}_{A}$ is converted back to $\mathbf{\hat{s}}_{B}$
using the inverse transformation that had converted it from $\mathbf{\hat{s}}_{B}$
to $\mathbf{\hat{s}}_{A}$.}
\label{fig:statetostate} 
\end{figure}

Liquid crystals in the nematic phase and which are parallel-aligned
have variable retardances, but a fixed rotation axis $\mathbf{\hat{k}}\left(2\Phi,0\right)$
in the $\mathbf{\hat{s}}_{1}$$\mathbf{\hat{s}}_{2}$ plane. Since
they have a fixed axis with linear polarization eigenstates, fewer
waveplates are necessary in Eq.~(\ref{eq:gen_rot_axis}): only a
quarter- and half-wave plate are required to transform any linear
polarization to an arbitrary polarization state \cite{Bhandari1989}.
Consequently, a four waveplate combination is sufficient for a liquid
crystal cell to effectively create an arbitrary rotation $\mathbf{R}_{\mathbf{k}_{B}(\varphi,\theta)}(\Delta)$,
\begin{eqnarray}
\mathbf{R}_{\mathbf{k}_{B}(\varphi,\theta)}(\Delta) & = & \mathbf{QWP}\left\llbracket \frac{\varphi}{2}+90^{\circ}\right\rrbracket \mathbf{HWP}\left\llbracket \frac{\varphi-\theta}{4}+\frac{\Phi}{2}+90^{\circ}\right\rrbracket \nonumber \\
 & \cdot & \mathbf{R}_{\mathbf{k}\left(2\Phi,0\right)}(\Delta)\mathbf{HWP}\textrm{\ensuremath{\left\llbracket \frac{\varphi-\theta}{4}+\frac{\Phi}{2}\right\rrbracket }}\mathbf{QWP}\textrm{\ensuremath{\left\llbracket \frac{\varphi}{2}\right\rrbracket }}.\label{eq:rot_axis}
\end{eqnarray}
The action of the waveplates in Eq.~(\ref{eq:rot_axis}) is shown
in Fig. \ref{fig:statetostate}(b) for $\mathbf{\hat{k}}\left(2\Phi,0\right)=\mathbf{\hat{s}}_{1}$.
We will repeatedly use this method of converting rotation axes on
the Poincaré sphere in the following sections.

\section{General polarization transformations with spatial light modulators
\label{sec:general_pol_transfo}}

\subsection{Spatially variable rotation about a fixed axis\label{subsec:Spatially-Variable-Rotation}}

If many liquid crystal cells are arranged in an array, we obtain a
spatial light modulator (LC-SLM). Here, For simplicity, we take the
fast axis of each liquid crystal cell, and thus the whole LC-SLM,
to be along the horizontal. i.e., with $\Phi=0$. If one used uniformly
H polarized light, a nematic-phase parallel aligned LC-SLM would function
as it is commonly used: as a ``phase-only'' spatial light modulator.
Here, we consider other input polarizations. The LC-SLM has a rotation
matrix equivalent to Eq.~(\ref{eq:X}) but with a spatially variable
retardance (or ``phase distribution'') of $\Delta(x,y)$. In this
way, LC-SLMs provide the spatial degree of freedom to extend the general
polarization transformation in Eq.~(\ref{eq:rot_axis}) to be, 
\begin{gather}
\mathbf{R}_{\mathbf{k}_{B}(\varphi,\theta)}(\Delta(x,y))=\nonumber \\
\mathbf{QWP}\left\llbracket \frac{\varphi}{2}+90^{\circ}\right\rrbracket \mathbf{HWP}\left\llbracket \frac{\varphi-\theta}{4}+90^{\circ}\right\rrbracket \mathbf{SLM}(\Delta(x,y))\mathbf{HWP}\textrm{\ensuremath{\left\llbracket \frac{\varphi-\theta}{4}\right\rrbracket }}\mathbf{QWP}\textrm{\ensuremath{\left\llbracket \frac{\varphi}{2}\right\rrbracket }}.\nonumber \\
\label{eq:SLM_axis_rot}
\end{gather}
The corresponding rotation matrix can then be computed by using Eq.~(\ref{eq:wp})
for the quarter- and half-wave plates, and Eq.~(\ref{eq:X}) for
the LC-SLM. This configuration gives the possibility for the whole
LC-SLM to have an arbitrary fixed rotation axis $\hat{\mathbf{k}}_{B}(\varphi,\theta)$,
but with a spatially variable rotation angle $\Delta(x,y)$. Fig.~\ref{fig:statetostate}(b)
traces the path of a state $\mathbf{\hat{s}}_{B}=\mathbf{\hat{k}}_{B}$
on the Poincaré Sphere as it travels through the waveplates in Eq.~(\ref{eq:SLM_axis_rot}).

\subsection{Practical examples and special cases}

Let us look at some practical examples of Eq. (\ref{eq:SLM_axis_rot}).

(\textit{i}) For a rotation axis on the equator (i.e., $\mathbf{\hat{s}}_{1}$$\mathbf{\hat{s}}_{2}$
plane), the quarter-wave plates in Eq.~(\ref{eq:SLM_axis_rot}) can
be removed, and $\varphi=0$. For example, for rotations about $\mathbf{\hat{k}}_{B}=\mathbf{\hat{s}}_{2}$,
the diagonal polarization axis, the sequence would be $\mathbf{R}_{\mathbf{s}_{2}}(\Delta(x,y))=\mathbf{HWP}\left\llbracket 112.5^{\circ}\right\rrbracket \mathbf{SLM}(\Delta(x,y))\mathbf{HWP}\textrm{\ensuremath{\left\llbracket 22.5^{\circ}\right\rrbracket }}$.

(\textit{ii}) For rotations about a circular polarization axis, i.e.,
the $\mathbf{\hat{s}}_{3}$ axis, we can remove the half-wave plates,
and use the sequence $\mathbf{R}_{\mathbf{s}_{3}}(\Delta(x,y))=\mathbf{QWP}\left\llbracket -45^{\circ}\right\rrbracket \mathbf{SLM}(\Delta(x,y))\mathbf{QWP}\textrm{\ensuremath{\llbracket45^{\circ}\rrbracket}}.$
This sequence has been demonstrated to be particularly useful in creating
vector beams such as radial and azimuthal polarization distributions
\cite{Zhuang2000}. For this, one begins with uniform horizontally
polarized light, $\mathbf{E}=\hat{\mathbf{x}}=\left(\hat{\mathbf{l}}+\hat{\mathbf{r}}\right)/\sqrt{2}$.
The right-circular component is retarded with respect to the left-circular
component resulting in the following polarization distribution after
the waveplates and LC-SLM, 
\begin{equation}
\mathbf{E}=\frac{1}{\sqrt{2}}\left(\hat{\mathbf{l}}+e^{i\Delta(x,y)}\hat{\mathbf{r}}\right)=\left(\cos\left(\frac{\Delta(x,y)}{2}\right)\hat{\mathbf{x}}-\sin\left(\frac{\Delta(x,y)}{2}\right)\hat{\mathbf{y}}\right)e^{i\Delta(x,y)/2}.\label{eq:radial}
\end{equation}

The polarization remains linear and is rotated in the laboratory by
an angle of $\Delta(x,y)$/2. Thus, the linear polarization direction
can vary spatially in an arbitrary manner. However, on the right side
of Eq.~(\ref{eq:radial}) there appears an additional phase term
that will not explicitly arise in our analysis using rotations in
the Poincaré sphere. This phase is addressed in the Appendix B.

\begin{figure}[ht!]
\centering\includegraphics[width=0.95\textwidth]{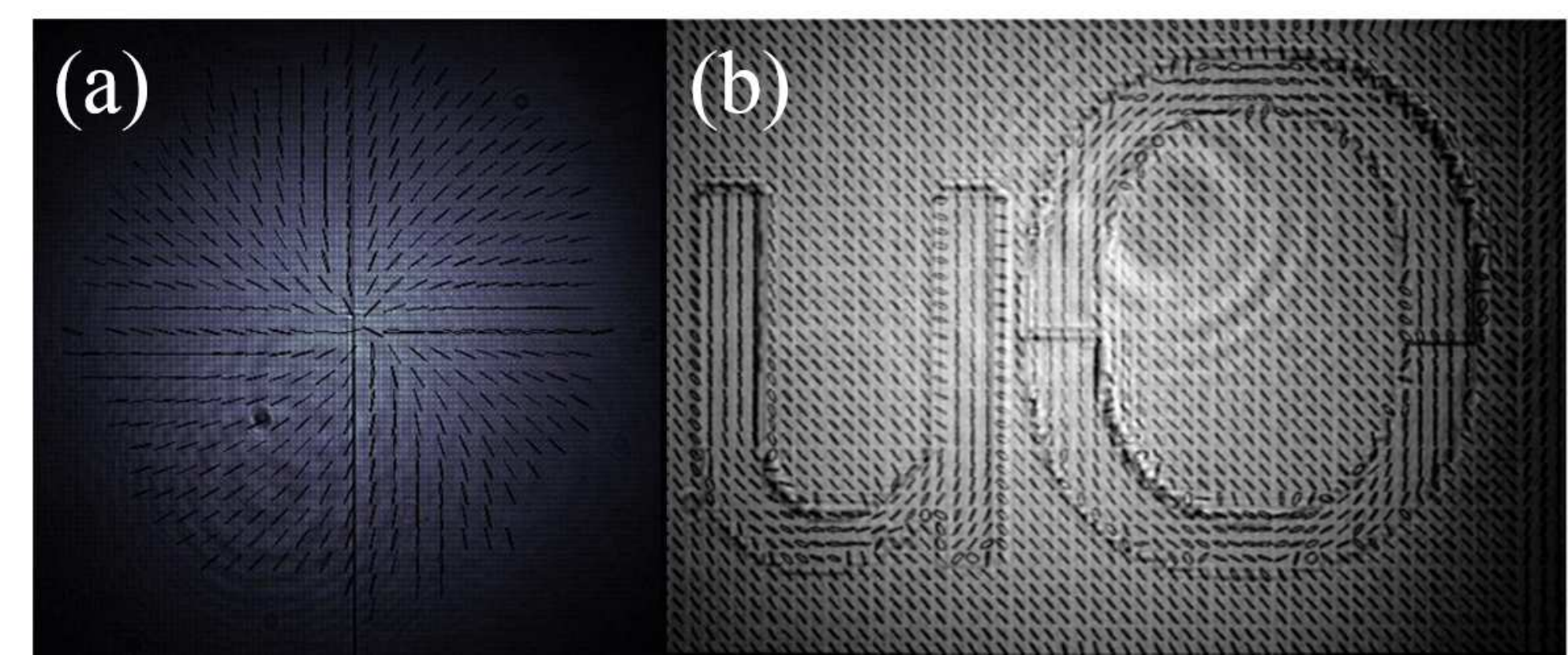} \protect\caption{Arbitrary linear polarization rotations. (a) Rotations about $\mathbf{\hat{s}}_{3}$
axis to produce a radial polarization distribution. The black lines
indicate the polarization direction, and the grey-scale gives the
intensity. (b) This technique can be used to write arbitrary linear
polarization patterns. Here, we write the initials of the University
of Ottawa with polarization. See Appendix B for experimental details.}
\label{fig:rad} 
\end{figure}

In Fig. \ref{fig:rad}(a) we demonstrate that the polarization can
be aligned radially. In Fig. \ref{fig:rad}(b), the polarization follows
the local asymptote of the letters ``uO'' (i.e., University of Ottawa).
The presence of light at the center may seem surprising since radially
polarized beams have a null in intensity at their center, as in a
Laguerre-Gauss mode beam. However, in Fig. \ref{fig:rad} we are imaging
the LC-SLM surface, at which only the phase, rather than intensity
is changed. The resulting field distribution is no longer a ``beam''
(for more detail on this subject see Appendix B).



\subsection{Transformation from an arbitrary polarization state to another arbitrary
polarization state\label{subsec:arb_arb}}

In our next step in increasing generality, we introduce a scheme to
transform an arbitrary input polarization $\mathbf{\hat{s}}_{i}$
to another arbitrary output polarization, $\mathbf{\hat{s}}_{o}$.
Both polarizations can spatially vary independently. Crucially, though,
both $\mathbf{\hat{s}}_{i}$ and $\mathbf{\hat{s}}_{o}$ must be known
at every position $\left(x,y\right)$ before implementing the transformation,
potentially through prior measurements.

To achieve this transformation, a second LC-SLM is added to the compound
device of Eq.~(\ref{eq:SLM_axis_rot}), thereby giving the ability
to perform rotations about two distinct axes on the Poincaré sphere.
Through two rotations about two orthogonal axes, say $\mathbf{\hat{s}}_{1}$
and $\mathbf{\hat{s}}_{2}$, one can transform any point to any other
point on the unit sphere \cite{Koks2006}. However, this only holds
if one can change the order of the orthogonal rotations, depending
on the input and output states. In contrast, we consider a fixed ordering
defined by the sequence of LC-SLMs and waveplates in the experimental
setup.

\begin{figure}[ht!]
\centering\includegraphics[width=9cm]{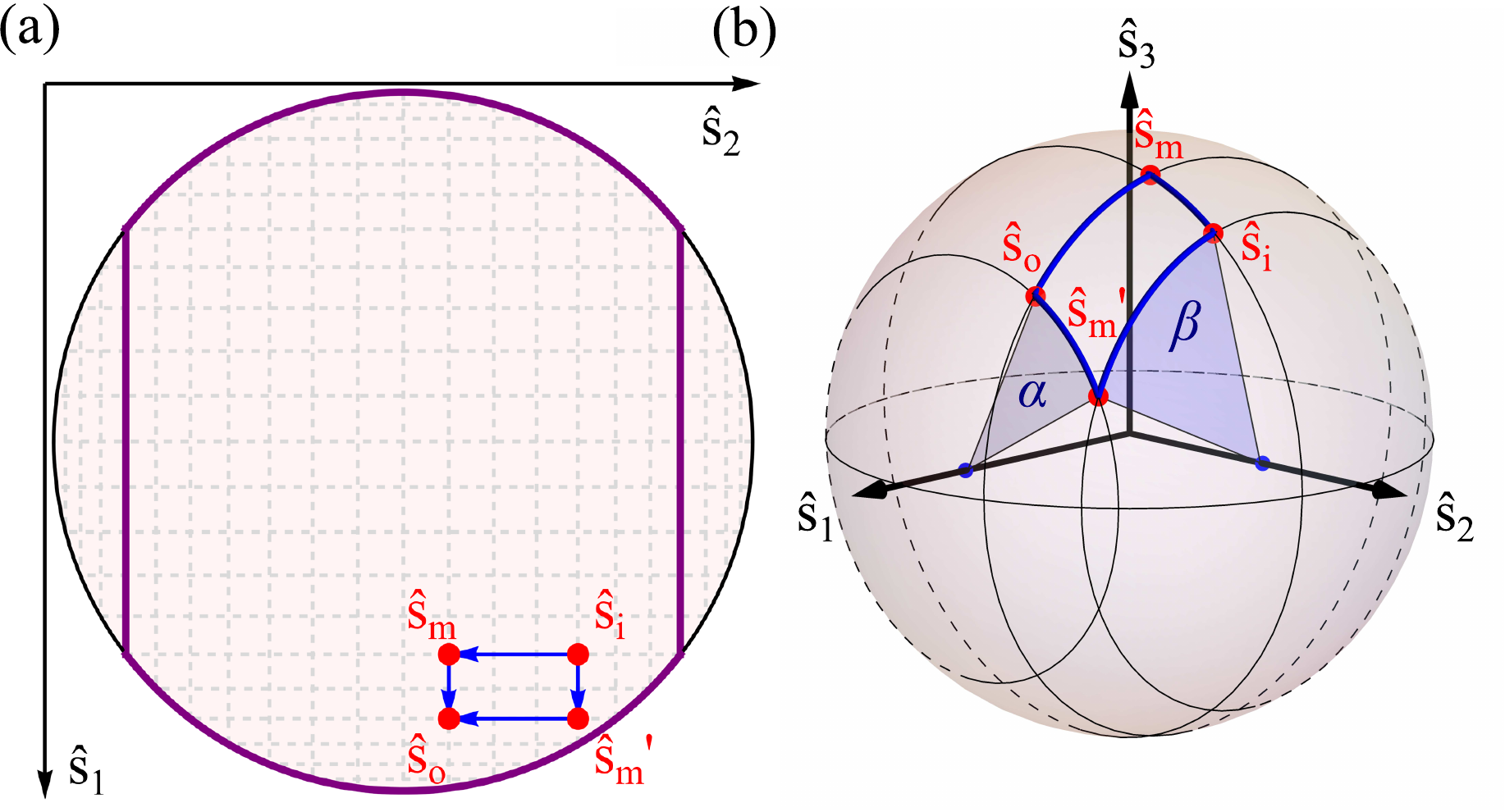} \protect\caption{Representation of the transformation of an arbitrary polarization
state $\mathbf{\hat{s}}_{i}$ into another arbitrary polarization
state $\mathbf{\hat{s}}_{o}$ using 2 LC-SLMs. (a) represents the
top view of the Poincaré Sphere depicted in (b). The input state $\mathbf{\hat{s}}_{i}$
can be converted to $\mathbf{\hat{s}}_{o}$ by a succession of one
rotation about $\mathbf{\hat{s}}_{1}$ and one rotation about $\mathbf{\hat{s}}_{2}$.
If the order of rotation is not defined, two intermediate points are
possible: $\mathbf{\hat{s}}_{m}$ if the state is first rotated about
$\mathbf{\hat{s}}_{1}$, and $\mathbf{\hat{s}}_{m'}$ if the state
is first rotated about $\mathbf{\hat{s}}_{2}$. However, if the rotation
about $\mathbf{\hat{s}}_{1}$ is the first one performed, then, there
is only one possibility to go from $\mathbf{\hat{s}}_{i}$ to $\mathbf{\hat{s}}_{o}$,
which means that it is impossible to reach any polarization state.
In this case, the accessible states are in the region outlined in
purple.}
\label{fig:2SLMcase} 
\end{figure}

In this section, it is useful to visualize the Poincaré sphere by
projecting it on a plane spanned by our two rotation axes, $\mathbf{\hat{s}}_{1}$
and $\mathbf{\hat{s}}_{2}$. This is shown in Fig. \ref{fig:2SLMcase}(a).
A rotation about $\mathbf{\hat{s}}_{1}$ will take $\mathbf{\hat{s}}_{i}$
to $\mathbf{\hat{s}}_{m}$. A following rotation about $\mathbf{\hat{s}}_{2}$
takes $\mathbf{\hat{s}}_{m}$ to $\mathbf{\hat{s}}_{o}$. Fig. \ref{fig:2SLMcase}(b)
shows this sequence in a 3-dimensional view of the Poincaré sphere
for reference. Reversing the order of the rotations would also work,
but would instead pass through the intermediate state $\mathbf{\hat{s}}_{m'}$.
This will not be generally true though. For many pairs of states $\mathbf{\hat{s}}_{i}$
and $\mathbf{\hat{s}}_{o}$, only one ordering will work. In particular,
the region outlined in purple in Fig.~\ref{fig:2SLMcase}(a) contains
the subset of output states $\{\mathbf{\hat{s}}_{o}\}$ that can be
reached from the specific $\mathbf{\hat{s}}_{i}$ when rotating about
$\mathbf{\hat{s}}_{1}$ first, followed by $\mathbf{\hat{s}}_{2}$.
To understand this, consider moving horizontally in either direction
from $\mathbf{\hat{s}}_{i}$ along a line at $s_{1}^{i}=\mathbf{\hat{s}}_{1}\cdot\mathbf{\hat{s}}_{i}$
(i.e., a rotation about $\mathbf{\hat{s}}_{1}$). In doing so, the
largest achievable magnitude for $s_{2}^{o}=\mathbf{\hat{s}}_{2}\cdot\mathbf{\hat{s}}_{o}$
is set by the intersection of the line with the circle bounding the
Poincaré sphere. That is, at $1=(s_{1}^{i})^{2}+(s_{2}^{o})^{2}$.

\subsubsection{Required retardances}

Keeping this restriction in mind, we now calculate the required retardances,
$\alpha$ and $\beta$, of the two LC-SLMs. We assume a configuration
which rotates first about $\mathbf{\hat{s}}_{1}$ (by $\alpha$) then
$\mathbf{\hat{s}}_{2}$ (by $\beta$): 
\begin{equation}
\mathbf{T}_{\mathbf{s}_{i}\rightarrow\mathbf{s}_{o}}=\mathbf{R}_{\mathbf{s}_{2}}(\beta)\mathbf{R}_{\mathbf{s}_{1}}(\alpha)=\mathbf{HWP}\llbracket112.5^{\circ}\rrbracket\mathbf{SLM}(\beta)\mathbf{HWP}\textrm{\ensuremath{\llbracket22.5^{\circ}\rrbracket\mathbf{SLM}(\alpha).}}\label{eq:arb_arb}
\end{equation}
While for brevity we omit an explicit spatial dependence in these
vectors and retardances, it should be understood to be implicit in
what follows. Specifically, all quantities are for the same transverse
point in the light field. In order to transform an input polarization
$\mathbf{\hat{s}}_{i}=[s_{1}^{i},s_{2}^{i},s_{3}^{i}]$ to a target
output polarization $\mathbf{\hat{s}}_{o}=[s_{1}^{o},s_{2}^{o},s_{3}^{o}]$,
one requires the following retardances: 

\begin{flalign}
\alpha'  = & \hspace{1mm} \text{atan2}\left(\left[0,s_{2}^{i},s_{3}^{i}\right]\cdot\left[0,s_{2}^{m},s_{3}^{m}\right],~\text{sign}(s_{2}^{i}s_{3}^{m}-s_{3}^{i}s_{2}^{m})\|\left[0,s_{2}^{i},s_{3}^{i}\right]\times\left[0,s_{2}^{m},s_{3}^{m}\right]\|\right),\quad\label{eq:phiX}\\
\beta'  = & \hspace{1mm}  \text{atan2}\left(\left[s_{1}^{m},0,s_{3}^{m}\right]\cdot\left[s_{1}^{o},0,s_{3}^{o}\right],~\text{sign}(s_{3}^{m}s_{1}^{o}-s_{1}^{m}s_{3}^{o})\|\left[s_{1}^{m},0,s_{3}^{m}\right]\times\left[s_{1}^{o},0,s_{3}^{o}\right]\|\right),\quad\label{eq:phiY}\\
\mathbf{\hat{s}}_{m}  = &  \left[s_{1}^{i},~s_{2}^{o},~\text{sign}(s_{3}^{o})\sqrt{(s_{3}^{i})^{2}+(s_{2}^{i})^{2}-(s_{2}^{o})^{2}}\right],\label{eq:mid} 
\end{flalign}
where $\mathbf{\hat{s}}_{m}=[s_{1}^{m},s_{2}^{m},s_{3}^{m}]$ is the
intermediate point, $\cdot$ is the dot product, $\times$ is the
vector cross product, $\|\hat{s}\|=\sqrt{s_{1}^{2}+s_{2}^{2}+s_{3}^{2}}$
is the vector norm, and $\mathrm{sign}$ is the standard signum function.
The function $\text{atan2}(x,y)$ is defined as the angle between
the positive $x$ axis and the point $(x,y)$, with angle increasing
towards the positive $y$ axis. Additionally, we take $\mathrm{\alpha=\alpha'}+2k\pi$
and $\mathrm{\beta=\beta'}+2k\pi$, in order to ensure that the two
LC-SLMs rotate in the positive sense (the angle is increasing). 

\subsubsection{Applications of arbitrary to arbitrary polarization transformations}

We now present three examples of applications that use this transformation.

\begin{figure}[ht!]
\centering\includegraphics[height=5cm]{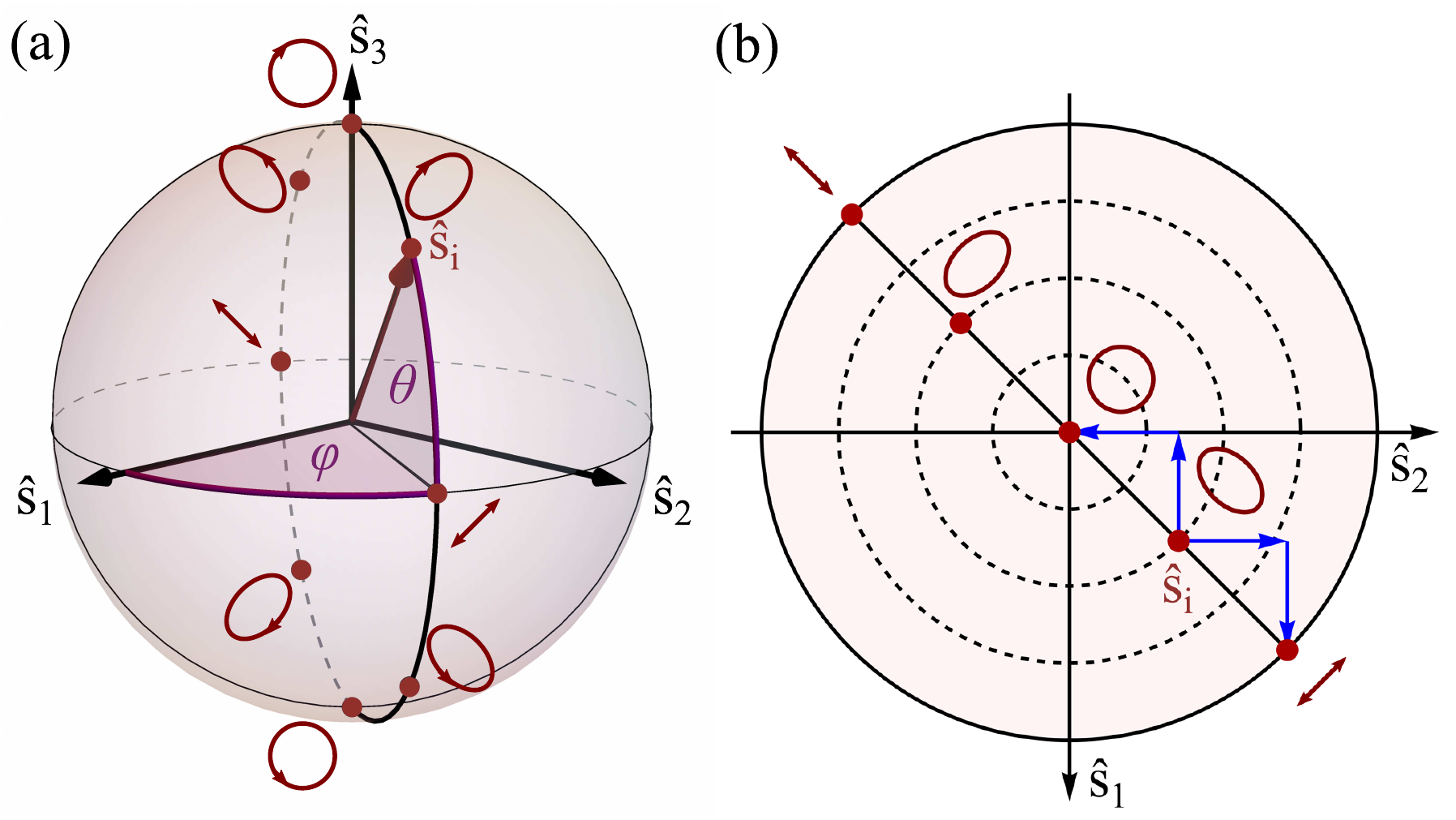} \protect\caption{Representation of the action of a beam healer / beam ellipticity changer
on the Poincaré Sphere. (b) represents the top view of the Poincaré
Sphere depicted in (a). Here, an elliptical polarization state $\mathbf{\hat{s}}_{i}$,
characterized by its coordinates $\phi$ and $\theta$, can be transformed
into any states laying on the great circle passing through the poles
and itself using two LC-SLMs. In particular, the state $\mathbf{\hat{s}}_{i}$
can be converted to a linear state, in which case its ellipticity
is removed, or converted into the state diametrically opposite on
the Poincaré Sphere, in which case the sign of its ellipticity is
flipped. }
\label{fig:beamhealing} 
\end{figure}

\textbf{Ellipticity (de)magnifier:} The ellipticity of an input state
can be either magnified to be more circular, or demagnified to be
more linear. In terms of the Poincaré sphere, changing the ellipticity
of state $\mathbf{\hat{s}}_{i}$ corresponds to changing the input
state's polar angle $\theta$, while maintaining its azimuthal angle
$\varphi$. This could be used to change a light field containing
a spatial polarization distribution with an assortment of elliptical
states to one with only linear states. It could also flip the polarization
handedness. An example of the rotation paths is shown in Fig.~(\ref{fig:beamhealing}).

\textbf{Beam healer:} Passage through birefringent optical media can
undesirably transform a uniform polarization into a non-uniform polarization
distribution. This effect could potentially degrade the performance
of imaging systems. The method previously introduced in this section
(i.e., Eq.~(\ref{eq:arb_arb})) can restore polarization uniformity.
However, in order for the method to work for every polarization in
the non-uniform distribution, the uniform output polarization must
be either right or left-handed circular. In short, the transformation
is $\mathbf{T}_{\mathbf{s}_{i}\rightarrow\pm\mathbf{s}_{3}}$.

While transforming an arbitrary polarization to another arbitrary
polarization might seem completely general, it is not. We clarify
this point in Appendix B.

\subsection{Spatially variable retardation of a spatially variable polarization
distribution\label{subsec:fully_universal}}

The most general polarization transformation is a rotation by an arbitrary
angle $\xi$ about an arbitrary axis $\mathbf{\hat{k}}$ in the Poincaré
sphere, $\mathbf{R}_{\mathbf{k}}(\xi)$. This corresponds to a retardation
of an arbitrary polarization state. It is a universal unitary transformation
for polarization. The axis $\mathbf{\hat{k}}$ is defined by two free
parameters, its spherical coordinates $(\varphi,\theta)$, and rotation
angle $\xi$ adds a third parameter. It follows that one needs at
least three control parameters in order to implement this general
transformation. One solution is to use three variable liquid crystals
and appropriate fixed waveplates. Considering LC-SLMs, this would
implement a different general transformation at each and every transverse
position in a light field.

\begin{figure}[ht!]
\centering\includegraphics[width=0.95\textwidth]{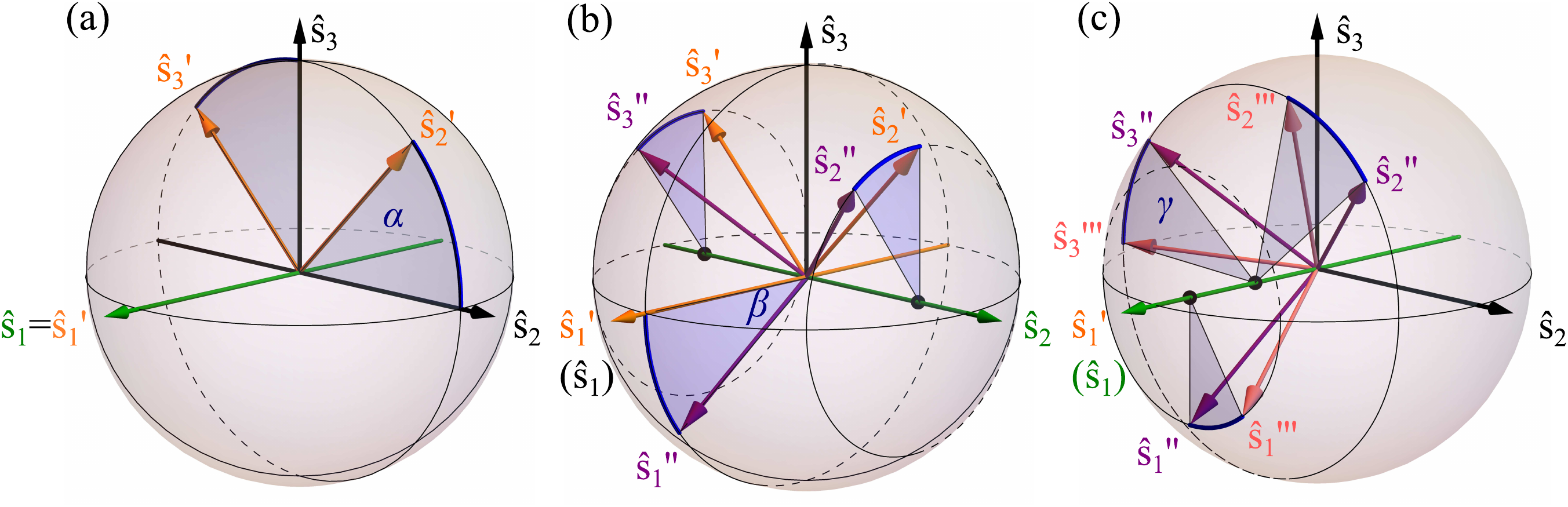} \protect\caption{Creation of a general unitary transformation using a sequence of three
orthogonal Euler rotations. The first rotation (a) is about $\mathbf{\hat{s}}_{1}$
by angle $\alpha$, the second (b) about $\mathbf{\hat{s}}_{2}$ by
$\beta$, and the third (c) about $\mathbf{\hat{s}}_{1}$ by $\gamma$.
From the transformation of the main axes of the Poincaré Sphere (e.g.,
$\mathbf{\hat{s}}_{1}\rightarrow\mathbf{\hat{s}}_{1}'\rightarrow\mathbf{\hat{s}}_{1}''\rightarrow\mathbf{\hat{s}}_{1}'''$)
one can see that this combination of rotations performs a general
3d rotation (e.g., yaw, pitch, and roll), and, consequently, a general
polarization rotation up to a global phase.}
\label{fig:3SLMcase3D} 
\end{figure}

As in section \ref{subsec:arb_arb}, the LC-SLMs and waveplates effectively
implement rotations about orthogonal axes. The addition of the third
LC-SLM, and thus third rotation, allows us to draw upon the concept
of proper extrinsic Euler angles, in which a general 3-d rotation
$\mathbf{R}_{\mathbf{k}}(\xi)$ can be decomposed into three successive
rotations about any two orthogonal axes. Here, we use $\mathbf{\hat{s}}_{1}$
and $\mathbf{\hat{s}}_{2}$. Accordingly, we compose the general rotation,

\begin{align}
\mathbf{\mathbf{R}_{\mathbf{k}}(\xi)}  = & \hspace{1mm} \mathbf{R}_{\mathbf{s}_{1}}(\gamma)\mathbf{R}_{\mathbf{s}_{2}}(\beta)\mathbf{R}_{\mathbf{s}_{1}}(\alpha) \nonumber \\ 
= & \hspace{1mm}  \mathbf{SLM}(\gamma)\mathbf{HWP}\llbracket112.5^{\circ}\rrbracket\mathbf{SLM}(\beta)\mathbf{HWP}\textrm{\ensuremath{\llbracket22.5^{\circ}\rrbracket\mathbf{SLM}(\alpha)}}\nonumber \\
  = &  \left[\begin{array}{ccc}
\cos\beta & \sin\alpha\sin\beta & \cos\alpha\sin\beta\\
\sin\beta\sin\gamma & \cos\alpha\cos\gamma-\cos\beta\sin\alpha\sin\gamma & -\cos\gamma\sin\alpha-\cos\alpha\beta\sin\gamma\\
-\cos\gamma\sin\beta & \cos\beta\cos\gamma\sin\alpha+\cos\alpha\sin\gamma & \cos\alpha\cos\beta\cos\gamma-\sin\alpha\sin\gamma
\end{array}\right].\nonumber \\
\end{align}
Fig. \ref{fig:3SLMcase3D} demonstrates the sequential rotations that
$\mathbf{R}_{\mathbf{k}}(\xi)$ performs on the principal axes in
terms of the three angles $\alpha$, $\beta$, and $\gamma$.

\subsubsection{Required retardances}

To simplify our notation we write $\mathbf{R\equiv R}_{\mathbf{k}}(\xi)$
in the following and define the $\mathbf{R}$ $_{ij}$ matrix element
to be the $i$ $^{th}$ row from the top and $j$ $^{th}$ column
from the left. In terms of these elements, the angle of each rotation
is,

\begin{eqnarray}
\alpha' & = & \text{atan2}(\mathbf{R}_{13},\mathbf{R}_{12}),\label{eq:alpha}\\
\beta' & = & \text{atan2}(\mathbf{R}_{11},\sqrt{\mathbf{R}_{21}^{2}+\mathbf{R}_{31}^{2}}),\label{eq:beta}\\
\gamma' & = & \text{atan2}(-\mathbf{R}_{31},\mathbf{R}_{21}).\label{eq:gamma}
\end{eqnarray}

These angles, modded by 2$\pi$, are the retardances, $\alpha,$ $\beta,$
and $\gamma$, that are applied at each transverse position in the
light field by the three LC-SLMs.

These angles are sufficient if one begins with the actual matrix for
$\mathbf{R}_{\mathbf{k}}(\xi)$, but it may be more useful if they
are expressed in terms of the rotation axis $\mathbf{\hat{k}}=[k_{1},~k_{1},~k_{3}]$
and angle $\xi$,

\begin{align}
\alpha'  = & \hspace{1mm}  \frac{k_{1}k_{2}(1-\cos\xi)-k_{3}\sin\xi}{k_{2}\sin\xi+k_{1}k_{3}(1-\cos\xi)},\\
\beta'  = & \hspace{1mm}  \frac{[(k_{3}\sin\xi+k_{1}k_{2}(1-\cos\xi))^{2}+(k_{1}k_{3}(1-\cos\xi)-k_{2}\sin\xi)^{2}]^{1/2}}{\cos\xi+k_{1}^{2}(1-\cos\xi)},\\
\gamma'  = & \hspace{1mm}  \frac{k_{3}\sin\xi+k_{1}k_{2}(1-\cos\xi)}{-(k_{1}k_{3}(1-\cos\xi)-k_{2}\sin\xi)}.
\end{align}
These follow from the matrix expression of the Euler-Rodrigues formula,
Eq.~(\ref{eq:E-R}) and Eq.~(\ref{eq:alpha})-(\ref{eq:gamma}).
With these retardances, a completely general polarization unitary
can be applied at each transverse position $\left(x,y\right)$ in
a light field.

\subsubsection{Applications and examples}

\textbf{Compensation of arbitrary spatially dependent birefringence:
}Now that we are able to implement a fully universal polarization
transformation, we can fully compensate for propagation through optical
media. For example, in propagation through a multi-mode optical fiber,
stresses and strains in the fiber typically create a small local birefringence
that effects supported modes differently. This leads to a spatially
dependent unitary $\mathbf{R\equiv R}_{\mathbf{k}}(\xi)$. If one
is attempting to use spatial and polarization multiplexing to communicate
over such a fiber, this unwanted transformation will cause cross-talk
and errors.

Spatially resolved polarization tomography can determine $\mathbf{R}_{\mathbf{k}(x,y)}\left(\xi(x,y)\right)$
for the fiber. Once this is known, the apparatus described in this
section could be placed after the fiber to implement $\mathbf{R}_{\mathbf{k}(x,y)}\left(-\xi(x,y)\right)$,
thereby undoing the transformation. Every spatial mode would emerge
with the same polarization that it had at the fiber input.

\section{Conclusion}

In order to completely control photons and unlock their full potential,
scientists must be able to arbitrarily manipulate all four degrees
of freedom that fully describe their state: time-frequency, the two
transverse position-momentum directions (e.g., x and y), and polarization.
In this paper, we described in detail methods to manipulate polarization
with liquid crystal devices in conjunction with fixed waveplates.
Most generally, we showed how to implement any possible polarization
transformation, a universal unitary. Since they are based on liquid
crystals, these unitaries can be varied or even completely reconfigured
in milliseconds and be computer controlled. Faster operation (e.g.,
sub-nanosecond) can be achieved by instead using an electro-optic
phase modulator \cite{Yariv2007}, which retards light in a similar
manner to a liquid crystal.

Combining these methods with spatial light modulators allows for novel
and broad control of the spatially varying polarizations of light-fields.
Beyond producing vector polarized beams, we proposed a number of applications
of these transformations including repairing polarization aberrations
in a beam, as a polarization magnifier, and for the compensation of
spatial-mode dependent birefringence in fiber-optic communications.
Given the broad use of polarization in industrial processes, commercial
products, and scientific research, we expect that these general polarization
methods will have many more applications in the near future. 

\section*{Funding}

This research was undertaken, in part, thanks to funding from the
Canada Research Chairs, NSERC Discovery, Canada Foundation for Innovation,
and the Canada Excellence Research Chairs program.

\section*{Appendix A: Polarization conventions and transformations \label{appendixA}}

In this section, we give a brief theoretical review of polarization
manipulation. The reader should be aware that there are many conflicting
conventions in use for polarization. This section presents a consistent
set of definitions (see Table \ref{tab:polconventions} for a summary)
with which to apply the schemes we introduce in the main text. We
characterize the polarization by the manner in which the electric
field oscillates. That is, by the normalized complex vector, 
\begin{equation}
\mathbf{E}=a_{x}~\hat{\mathbf{x}}+a_{y}e^{i\delta}~\hat{\mathbf{y}},\label{eq:pol_state}
\end{equation}
where $a_{x}$ and $a_{y}$ are the amplitudes of, and $\delta$ the
relative phase between, the \textit{x} and \textit{y} electric field
components, respectively, with $a_{x}^{2}+a_{y}^{2}=1$. With $\hat{\mathbf{z}}$
pointing in the direction of propagation, we use a right-handed co-ordinate
system. We follow the convention in Ref. \cite{saleh2013fundamentals}
by defining right-handed polarization to be clockwise rotating, as
seen by the receiver. When $\mathrm{\delta}\neq k\pi/2$, where $k$
is an integer, the polarization state is ``elliptical'' since the
electric field vector traces out an ellipse as a function of time.

\begin{figure}[ht!]
\centering\includegraphics[width=10cm]{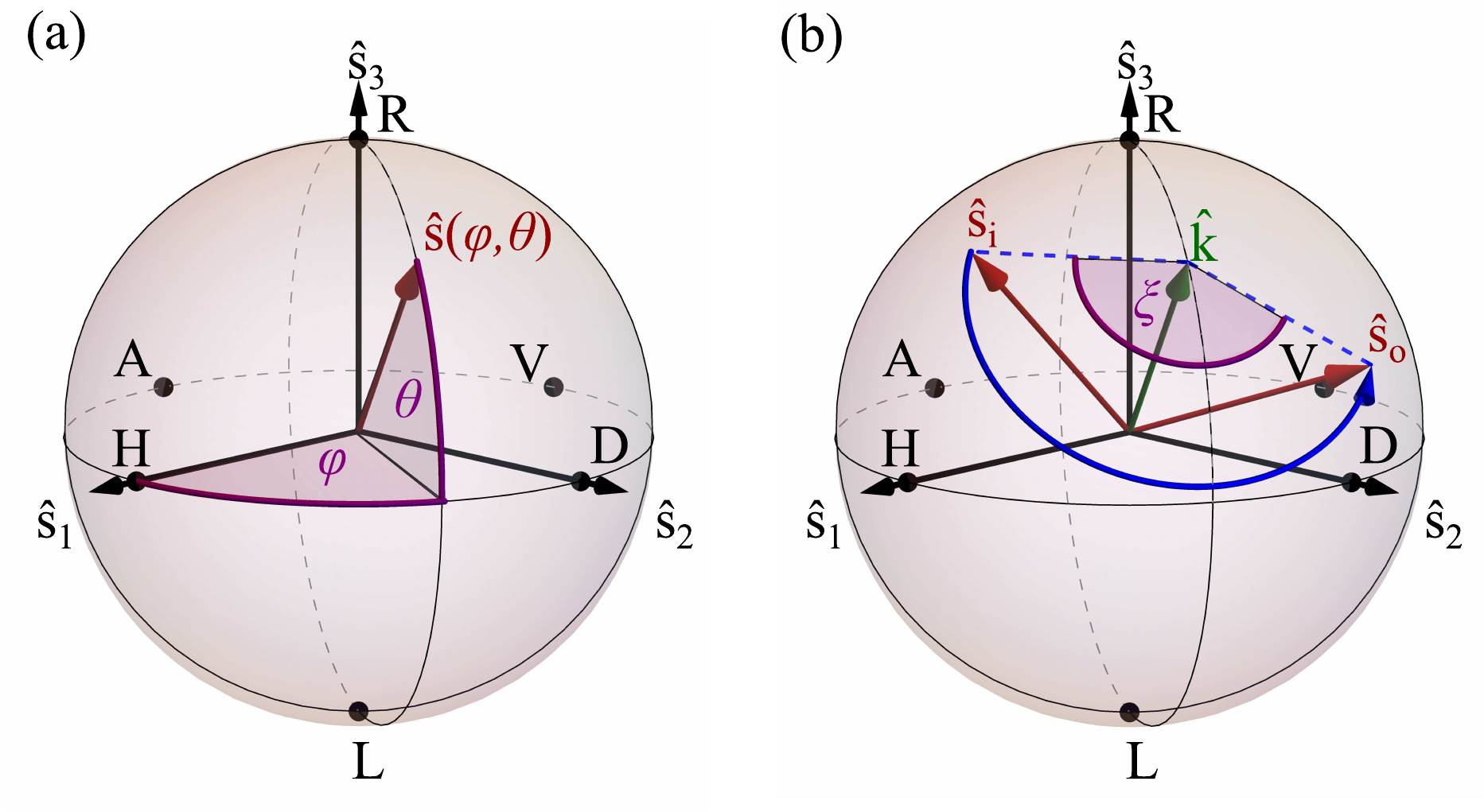} \protect\caption{The Poincaré sphere and polarization transformations. Polarization
states lie on the surface of a sphere that has a radius of one. Horizontal
(H) and vertical (V) polarizations define the $\mathbf{\hat{s}}_{1}$
axis, diagonal (D) and anti-diagonal (A) states define the $\mathbf{\hat{s}}_{2}$
axis, and left- (L) and right-hand (R) circular define the polar $\mathbf{\hat{s}}_{3}$
axis; each pair lie on the positive and negative ends of the axis,
respectively. (a) The polarization state in Eq.~(\ref{eq:pol_state})
is given by point $\mathbf{\hat{s}}\left(\varphi,\theta\right)=[s_{1},s_{2},s_{3}]$
on the surface, where $s_{i}$ are the Stokes parameters. The polarization
state $\mathbf{\hat{s}}$ can also be expressed by its coordinates
in the spherical system $\mathbf{\hat{s}}(\varphi,\theta)$. (b) In
a polarization transformation, any polarization state is rotated about
a fixed axis $\mathbf{\hat{k}}$ by an angle $\xi.$ \textit{Note:
Throughout this paper, the states are represented in red, the axes
of rotation in green, the transformation in blue and the definition
of angle in purple.}}
\label{fig:PS} 
\end{figure}

A visually intuitive representation for polarization states is to
represent them as points on the surface of a unit sphere, known as
the Poincaré sphere~\cite{Born:80}, as shown in Fig.~\ref{fig:PS}(a),
the polarization equivalent of the Bloch sphere for spin-1/2 or other
two-level systems. From the complex vector notation, we can calculate
the reduced (normalized) Stokes parameters~\cite{Born:80} to obtain
a polarization state's position on the sphere, 
\begin{eqnarray}
s_{0} & = & E_{x}E_{x}^{*}+E_{y}E_{y}^{*}=a_{x}^{2}+a_{y}^{2}=1,\\
s_{1} & = & E_{x}E_{x}^{*}-E_{y}E_{y}^{*}=a_{x}^{2}-a_{y}^{2}=\cos\varphi\cos\theta,\\
s_{2} & = & E_{x}E_{y}^{*}+E_{y}E_{x}^{*}=2a_{x}^{2}a_{y}^{2}\cos\delta=\sin\varphi\cos\theta,\\
s_{3} & = & i(E_{x}E_{y}^{*}-E_{y}E_{x}^{*})=2a_{x}^{2}a_{y}^{2}\sin\delta=\sin\theta,\quad
\end{eqnarray}
where $\varphi$ and $\theta$ are the azimuthal and polar angles
of the Poincaré sphere. Here, we consider left- and right-handed circular
polarizations are respectively mapped to the north and south poles
of the sphere. Linear polarization states lie along the equator and
elliptical states everywhere else. Orthogonal polarization states
are diametrically opposed points. The six polarizations that define
the axes are listed in Table 1.

\begin{table}[h]
\begin{centering}
\begin{tabular}{|c|c|c|c|c|c|}
\hline 
\multicolumn{2}{|c|}{} & \multicolumn{2}{c|}{Laboratory } & \multicolumn{2}{c|}{Poincaré Sphere}\tabularnewline
\hline 
Polarization state  & Abbrv.  & Vector  & Ellipse  & Stokes vector $\mathbf{\hat{s}}$  & $(\varphi,\theta)$ \tabularnewline
\hline 
\hline 
Horizontal  & H  & $\hat{\mathbf{x}}$  & $-$  & $[1,0,0]$  & $(0,0)$ \tabularnewline
\hline 
Vertical  & V  & $\hat{\mathbf{y}}$  & $|$  & $[-1,0,0]$  & $(\pi,0)$ \tabularnewline
\hline 
Diagonal  & D  & $\hat{\mathbf{d}}=\frac{\left(\hat{\mathbf{x}}+\hat{\mathbf{y}}\right)}{\sqrt{2}}$  & $\diagup$  & $[0,1,0]$  & $(\pi/2,0)$ \tabularnewline
\hline 
Anti-Diagonal  & A  & $\hat{\mathbf{a}}=\frac{\left(\hat{\mathbf{x}}-\hat{\mathbf{y}}\right)}{\sqrt{2}}$  & $\diagdown$  & $[0,-1,0]$  & $(-\pi/2,0)$\tabularnewline
\hline 
Right-Hand Circular  & R  & $\hat{\mathbf{r}}=\frac{\left(\hat{\mathbf{x}}+i\hat{\mathbf{y}}\right)}{\sqrt{2}}$  & $\circlearrowright$  & $[0,0,1]$  & $(0,\pi/2)$ \tabularnewline
\hline 
Left-Hand Circular  & L  & $\hat{\mathbf{l}}=\frac{\left(\hat{\mathbf{x}}-i\hat{\mathbf{y}}\right)}{\sqrt{2}}$  & $\circlearrowleft$  & $[0,0,-1]$  & $(0,-\pi/2)$ \tabularnewline
\hline 
\end{tabular}
\par\end{centering}
\caption{Table of the conventions used through this paper to define the principal
polarization states (i.e the main axes of the Poincaré sphere), their
abbreviations and their representations both in the laboratory frame
and on the Poincaré sphere. For the pictorial representation in the
Ellipse column, the direction of propagation is towards the observer.}
\label{tab:polconventions} 
\end{table}

In this paper, we consider only completely polarized states of light.
Consequently, the length of the reduced Stokes vector, defined by
$s_{0}$, is identically one, such that all states lie on the surface
of the Poincaré sphere. Henceforth, we drop $s_{0}$ so that a polarization
state is described by a three element Stokes vector $\mathbf{\hat{s}}=s_{1}\mathbf{\hat{s}}_{1}+s_{2}\mathbf{\hat{s}}_{2}+s_{3}\mathbf{\hat{s}}_{3}=[s_{1},s_{2},s_{3}]$.
Alternately, this unit vector can be equivalently expressed in spherical
coordinates as $\mathbf{\hat{s}}(\varphi,\theta)$. The angles $\varphi$
and $\theta$ can be related to the complex vector notation via, 
\begin{eqnarray}
\sin\varphi & = & 2a_{x}^{2}a_{y}^{2}\sin\delta,\\
\tan\theta & = & \frac{2a_{x}^{2}a_{y}^{2}\sin\delta}{\sqrt{(a_{x}^{2}-a_{y}^{2})^{2}+4a_{x}^{4}a_{y}^{4}\cos^{2}\delta}}.
\end{eqnarray}

In terms of these parameters, we use the conventions listed in Table
\ref{tab:polconventions}.

As shown in Fig.~\ref{fig:PS}(b), on the Poincaré sphere, polarization
transformations \textemdash{} general polarization unitaries \textemdash{}
are right-handed rotations of a state $\mathbf{\hat{s}}$ about a
unit-length axis $\mathbf{\hat{k}}=[k_{1},k_{2},k_{2}]=\mathbf{\hat{k}}\left(\varphi,\theta\right)$
by an angle $\xi$. Mathematically, to perform such a rotation on
a Stokes vector, we use a standard three-dimensional \emph{active}
rotation matrix, $\mathbf{R}_{\mathbf{k}}(\xi)$. These $3\times3$
matrices are simply the lower-right sub-matrix of the $4\times4$
Mueller rotation matrices used commonly in polarization theory. Using
this matrix and writing the Stokes vector as a column, the rotated
vector is $\mathbf{\hat{s}}'=\mathbf{R}_{\mathbf{k}}(\xi)\mathbf{\hat{s}}$.
The general rotation matrix $\mathbf{R}_{\mathbf{k}}(\xi)$ is given
by a form of the Euler-Rodrigues formula \cite{Koks2006}, 
\begin{equation}
\mathbf{R}_{\mathbf{k}}(\xi)=\mathbf{I}+\sin\xi\mathbf{K}+(1-\cos\xi)\mathbf{K}^{2},\label{eq:E-R}
\end{equation}
where $\mathbf{I}$ is the $3\times3$ identity matrix, and $\mathbf{K}$
is the cross-product operation matrix of $\mathbf{\hat{k}}$, 
\begin{equation}
\mathbf{\hat{k}}\times=\mathbf{K}=\left[\begin{array}{ccc}
0 & -k_{3} & k_{2}\\
k_{3} & 0 & -k_{1}\\
-k_{2} & k_{1} & 0
\end{array}\right].
\end{equation}
This gives a rotation matrix of, 

\begin{equation*}
\mathbf{R}_{\mathbf{k}}(\xi)=
\end{equation*}
\begin{align}
\left[\begin{array}{ccc}
\cos\xi+k_{1}^{2}(1-\cos\xi) & k_{1}k_{2}(1-\cos\xi)-k_{3}\sin\xi & k_{1}k_{3}(1-\cos\xi)+k_{2}\sin\xi\\
k_{1}k_{2}(1-\cos\xi)+k_{3}\sin\xi & \cos\xi+k_{2}^{2}(1-\cos\xi) & k_{2}k_{3}(1-\cos\xi)-k_{1}\sin\xi\\
k_{1}k_{3}(1-\cos\xi)-k_{2}\sin\xi & k_{2}k_{3}(1-\cos\xi)+k_{1}\sin\xi & \cos\xi+k_{3}^{2}(1-\cos\xi)
\end{array}\right].\label{eq:E-R2}
\end{align}

The rotation matrices for rotations about the $\mathbf{\hat{s}}_{1}$,
$\mathbf{\hat{s}}_{2}$, and $\mathbf{\hat{s}}_{3}$ axes can thus
be computed from Eq.~(\ref{eq:E-R2}) using $\mathbf{\hat{k}}=[1,0,0]$,
$[0,1,0]$, and $[0,0,1]$, respectively, 
\begin{eqnarray}
\mathbf{R}_{\mathbf{s}_{1}}(\alpha) & = & \left[\begin{array}{ccc}
1 & 0 & 0\\
0 & \cos\alpha & -\sin\alpha\\
0 & \sin\alpha & \cos\alpha
\end{array}\right],\label{eq:X}\\
\mathbf{R}_{\mathbf{s}_{2}}(\beta) & = & \left[\begin{array}{ccc}
\cos\beta & 0 & \sin\beta\\
0 & 1 & 0\\
-\sin\beta & 0 & \cos\beta
\end{array}\right],\label{eq:Y}\\
\mathbf{R}_{\mathbf{s}_{3}}(\gamma) & = & \left[\begin{array}{ccc}
\cos\gamma & -\sin\gamma & 0\\
\sin\gamma & \cos\gamma & 0\\
0 & 0 & 1
\end{array}\right].\label{eq:Z}
\end{eqnarray}


\section*{Appendix B: Residual Spatial Phase, Relay Imaging, and non-Universal
Transformations \label{appendixB}}


In this paper, we concern ourselves only with the the spatial polarization
distribution of the beam and neglect spatially varying phases. Nonetheless,
the two are linked. There are three physical phases, $\delta$, $b$,
and $c$, in a single frequency paraxial optical field, $\mathbf{E}(x,y)=\left(a_{x}~\hat{\mathbf{x}}+a_{y}e^{i\delta(x,y)}~\hat{\mathbf{y}}\right)e^{ib(x,y)}e^{ic}$.
Phase $c$ is an overall global phase that is constant across $x$
and $y$, and, hence, not relevant for this paper (it is relevant
for interference with ancillary fields, as in an interferometer).
Phase $b(x,y)$ is between fields at different positions. Phase $\delta(x,y)$
is between field polarization components at $\left(x,y\right)$ and
can vary with $x$ and $y$. It is this last phase, as well as the
magnitude of each polarization component, that we manipulate in this
paper.


\subsection*{Residual Phase}

\begin{figure}[ht!]
\centering\includegraphics[width=13cm]{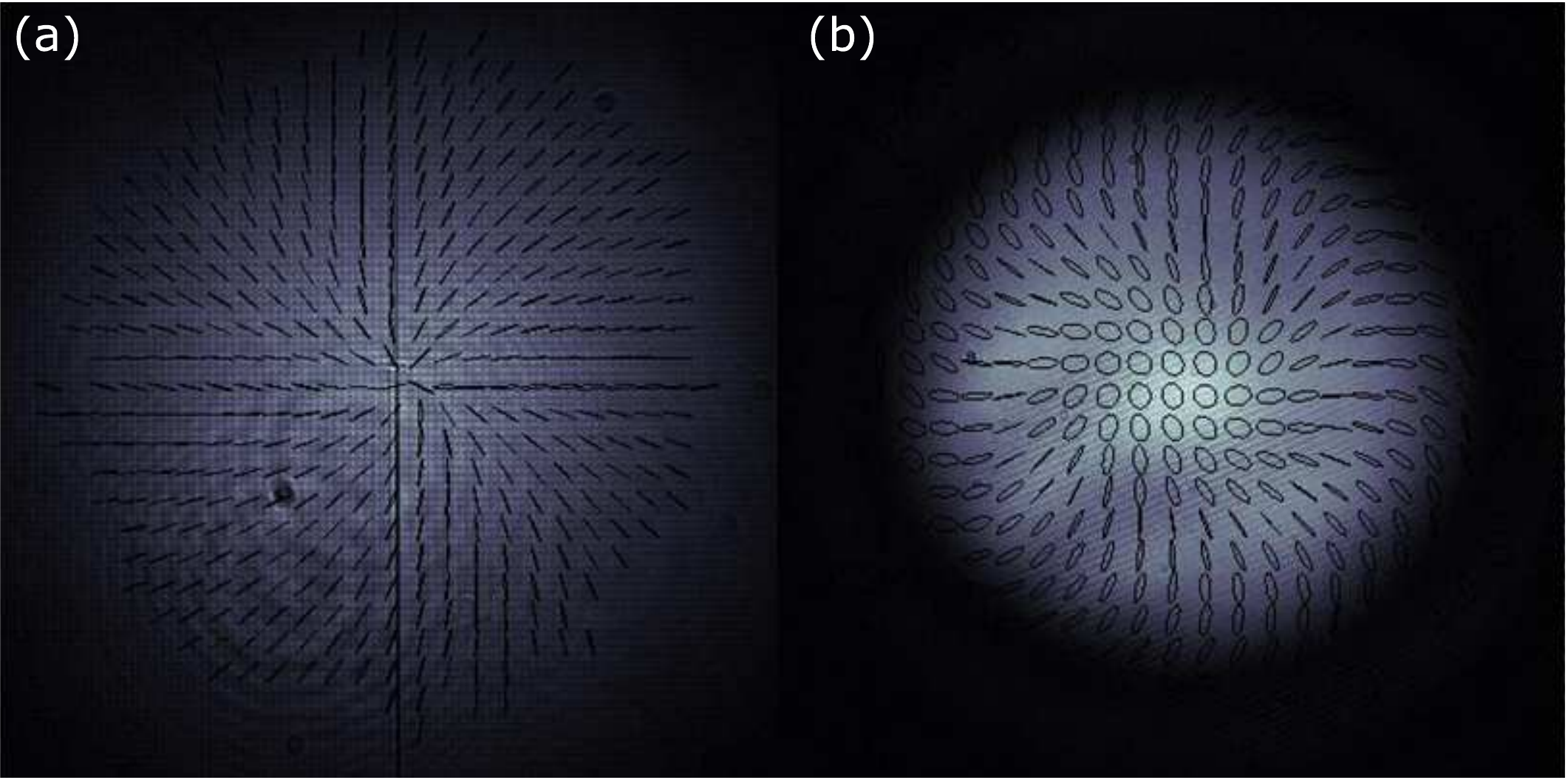} \protect\caption{Rotations about $\mathbf{\hat{s}}_{3}$ axis to produce radial polarization
distribution. We can see the pictures of the beam and its polarization
state at various positions in the imaging plane (a) and \char`\"{}far-field\char`\"{}
(b). }
\label{fig:rad2} 
\end{figure}

However, a residual spatially-varying phase $b\left(x,y\right)$ can
indeed arise in the polarization transformations that we present.
This phase is not evident from rotations in the Poincaré sphere but
does arise when using Jones Matrices \cite{Jones1947}. At the expense
of added complexity, one could compensate for $b(x,y)$ by adding
one additional LC-SLM to each of the transformations presented. If
not compensated, this residual phase can have physical consequences.
As an example, consider the case where we produce a radially polarized
field according to Eq. (\ref{eq:radial}). On the right-hand side,
a residual phase of $e^{i\Delta(x,y)}$ appears. Here, $\Delta(x,y)=2\phi$,
where $\phi$ is the azimuthal angle about the center of the radial
field. It's impact can be understood by considering the left-hand
side of Eq. (\ref{eq:radial}), $\mathbf{E}=\frac{1}{\sqrt{2}}\left(\hat{\mathbf{l}}+e^{i2\phi}\hat{\mathbf{r}}\right)$.
If an incoming optical field had a Gaussian transverse profile, the
left-handed component $\hat{\mathbf{l}}$ would be unchanged, whereas
the right-handed component $\hat{\mathbf{r}}$ would receive an azimuthally-varying
phase carrying $\ell=2$ units of orbital angular momentum (OAM).
This phase profile causes the right-handed light-field to no longer
be a solution to the paraxial wave equation. Hence, it is no longer
a ``beam'' in the sense that it does not maintain its spatial and
polarization distribution upon propagation (up to a scale factor).
We experimentally demonstrated this by allowing the radial field in
Fig.~\ref{fig:rad2}(a) to propagate 10 cm further. The polarization
and intensity distribution at this point is shown in Fig.~\ref{fig:rad2}(b).
We call this the ``far-field''.

\subsection*{Imaging requirements for general polarization manipulation}

Fig.~\ref{fig:rad} shows there are notable differences between the
output polarization distribution at the image plane of the LC-SLM
(near-field) and after propagation (far-field) for the same input
polarization and LC-SLM phase pattern, $\Delta(x,y)$. However, throughout
this paper we assumed that LC-SLMs can be placed in series while acting
on the same unchanged optical field. To achieve this, the field at
one LC-SLM is imaged onto the next with a 4-f system. The latter acts
as a relay imaging system with a one-to-one magnification ratio.

\begin{figure}[ht!]
\centering\includegraphics[width=0.5\textwidth]{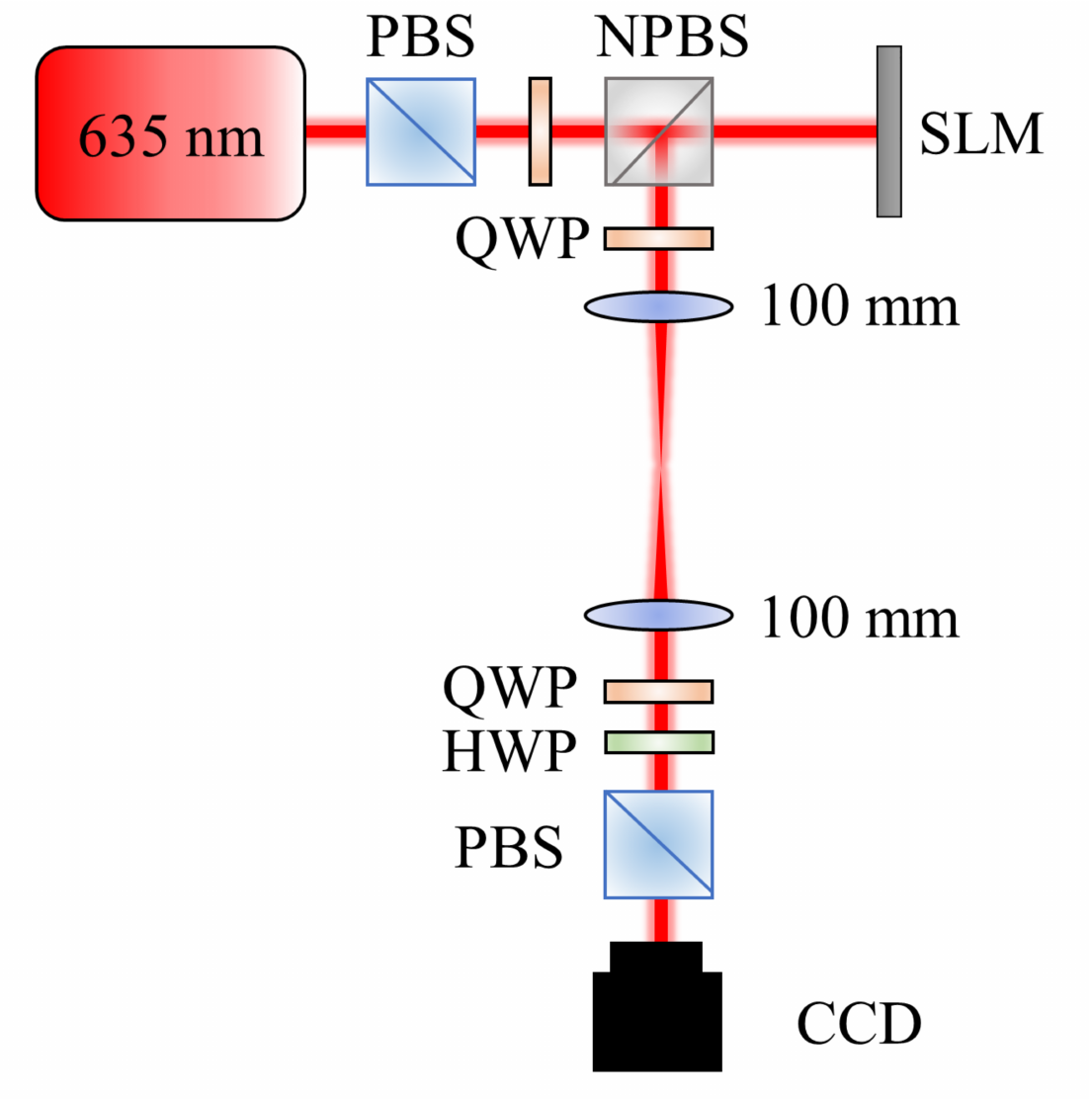} \protect\caption{Experimental setup to perform a rotation about $\mathbf{\hat{s}}_{3}$.
A diode laser (635~nm) is prepared to be left-handed circular with
a quarter-wave plate (QWP), and then reflected off of an LC-SLM (LCOS-SLM
X10468, Hamamatsu, Japan), which imprints the desired spatially varying
phase (i.e., rotation angle or ``retardance''). A non-polarizing
beam-splitter (NPBS) splits off half of the light to be analyzed.
A second quarter-wave plate converts circular states back to linear
states. A 4-f imaging system is used to image the plane of the LC-SLM
onto a CCD camera. We use short focal length (f = 100~mm, diameter
= 25.4~mm) doublet lenses in order to have a high numerical aperture.
The polarization of the light is then determined pixel by pixel via
polarization tomography (i.e., Stokes polarimetry) with a half-wave
plate (HWP), quater-wave plate, and polarizing beam-splitter (PBS)~\cite{Cardano:12}.}
\label{fig:expsetup} 
\end{figure}

Ideally, in order to create a simple setup with high transmissivity,
the LC-SLMs would be transmissive and placed in a line. However, reflective
LC-SLMs often have better performance specifications. A reflective
setup usually involves picking off a beam that is reflected at a small
angle. Since, the pick-off mirror acts as an aperture this can dramatically
change the optical field imaged by a 4-f setup. Our setup, shown in
Fig. \ref{fig:expsetup}, uses a reflective LC-SLM. In order to separate
the reflected beam from incident beam we use a non-polarizing beam-splitter
(NPBS) instead of a pick-off mirror. The drawback is that this NPBS
introduces loss. Nonetheless, the setup allows us to test the transformations,
which are lossless in principle.

\subsection*{Arbitrary to arbitrary polarization versus universal transformations}

While transforming an arbitrary polarization to another arbitrary
polarization might seem completely general, it is not. The most general
rotation is $\mathbf{R}_{\mathbf{k}}(\xi)$, whereas the transformation
that is implemented through this method is $\mathbf{R}_{\mathbf{s}_{2}}(\beta)\mathbf{R}_{\mathbf{s}_{1}}(\alpha)$.
The latter implements $\mathbf{T}_{\mathbf{s}_{i}\rightarrow\mathbf{s}_{o}}$,
which takes $\mathbf{\hat{s}}_{i}\rightarrow\mathbf{\hat{s}}_{o}$.
It also links the polarization states diametrically opposed to these
on the Poincaré sphere, $-\mathbf{\hat{s}}_{i}\rightarrow-\mathbf{\hat{s}}_{o}$.
However, $\mathbf{T}_{\mathbf{s}_{i}\rightarrow\mathbf{s}_{o}}$ does
not fix the retardance $\zeta$ between $\mathbf{\hat{s}}_{o}$ and
$-\mathbf{\hat{s}}_{o}$. This retardance is crucial when considering
how $\mathbf{T}_{\mathbf{s}_{i}\rightarrow\mathbf{s}_{o}}$ transforms
any input state other than $\pm\mathbf{\hat{s}}_{i}$. Polarizations
that are a superposition of $-\mathbf{\hat{s}}_{o}$ and $+\mathbf{\hat{s}}_{o}$
will transform in an undetermined way.

\section*{References}

\bibliographystyle{iopart-num.bst} 

\global\long\def\enquote#1{``#1''}

\end{document}